\documentstyle[11pt, amssymb]{article}
 \newcommand {\nc}{\newcommand}
 \nc{\eq}{\begin{equation}}
 \nc{\en}{\end{equation}}
 \nc{\eqa}{\begin{eqnarray}}
 \nc{\ena}{\end{eqnarray}}
 \def\nat{{\mathbb N}}
 \def\intg{{\mathbb Z}}
 \def\complex{{\mathbb C}}
 \def\Alg{{\mathcal A}}
 
 \def\Hs{{\mathcal H}}

 \def\inv{\j}
 \def\J{{\mathcal J}}
 \def\comp{{\mathcal K}}

 \def\tld{\tilde}
 \def\pr{\prime}
 \def\ot{\otimes}
 \def\dlt{\delta}
 \def\etal{{\it et al} }
 \newtheorem{definition}{Definition}
 
 \newtheorem{theorem}{Theorem}
 \newtheorem{proposition}{Proposition}
 \newtheorem{corollary}{Corollary}

 \def\prf{{\bf{Proof:}}\\}
 \def\endprf{$\blacksquare$}
 \title{Canonical Differential Calculi For Finitely Generated Abelian Group and Their Fermionic Representations}
 \author{Jian Dai, Xing-Chang Song\\
 Theory Group, Department of Physics, Peking University\\
 Beijing, P. R. China, 100871\\
 jdai@mail.phy.pku.edu.cn,
 songxc@ibm320h.phy.pku.edu.cn}
 \date{May 8th, 2001}
 
\begin{document}
%  \begin{titlepage}
  \maketitle
  \begin{abstract}
   Canonical differential calculus is defined for finitely generated abelian
   group with an involution existing consistently. Two such
   canonical calculi are found out. Fermionic representation for
   canonical calculus is defined based on quantized calculus.
   Fermionic representations for above-mentioned two canonical
   calculi are searched.


  \end{abstract}
%  \end{titlepage}
  \section{Introduction}
   Noncommutative differential calculus can be formulated on finite
   groups and some simple types of discrete infinite groups within
   the quantum-algebraic approach towards noncommutative geometry (NCG) \cite{S1}\cite{B1}. There are two problems which
   are far from being fully understood however:
   1) The choice of differential calculus is considerably arbitrary, even
   though translation invariance being included; and the arbitrariness corresponds to the power set
   of the group.
   2) Whether the concept of spinor can be generalized to this differential
   construction is unknown.\\

   These two problems for the category of
   {\it finitely generated abelian group} (FGAG) whose structures have
   been totally clear in group theory will be addressed here. {\it Canonical
   differential calculi} and their fermionic representations will be considered for these groups, where
   the term ``canonical'' refers to that the choice of the calculus
   reflects the groups structure properly and is compatible with group involution.
   Note that the first question for FGAGs has been
   considered in \cite{G1} in a less rigid sense.
   This article is organized as following. In Sec.\ref{secP},
   necessary mathematics is prepared, including
   differential calculus over finitely generated groups, classification
   of FGAGs. In
   Sec.\ref{secC}, canonical calculus is defined and two canonical
   calculi are found for FGAGs.
   In Sec.\ref{secF}, fermionic representation for canonical calculi is discussed and
   fermionic representations for the above two canonical calculi are searched. In Sec.\ref{secA},
   some open discussions are given.
  \section{Preliminaries and Notations}\label{secP}
   Contents in this section are well-known, and they are rephrased here just for integrality of this article.
   Group $G$ is required to be {\it finitely generated} and
   the term ``discrete'' is avoided due to the subtlety in topology.
   Dual of $\complex G$ is written as $\Alg[G]$, which becomes a commutative $\complex$-algebra
   under pointwise product.
   Induced right(left) actions of $G$ on $\Alg[G]$ are written as $R_g(L_g)$ for all $g$ in
   $G$. $e$ is the unit of $G$ as usual.
   A first order differential over $\Alg[G]$ is a pair $(M, d)$, where $M$ being an $\Alg[G]$-bimodule
   generated by a {\it left invariant basis} $\{\xi^g:g\in G^\pr\subset G\setminus
   \{e\}\}$, whose structure is characterized by the equation
   \eq\label{FN}
    \xi^g f=R_g(f)\xi^g, \forall f\in \Alg[G], g\in G^\pr
   \en
   and $d\in Hom_\complex(\Alg[G],M)$ fulfills Leibnitz rule.
   Eq.(\ref{FN}) is crucially important for first order differential on $G$
   in the sense that $\Alg[G]$ is a set-theoretical object on which no group structure is encoded,
   therefore this {\it noncommutativity} endows group structure
   onto $\Alg[G]$, providing $G^\pr$ generates $G$.
   Generically for all $f$ in $\Alg[G]$, $d(f)=\partial_g(f)\xi^g$ with
   $\partial_g(f)\in\Alg[G]$;
   accordingly,
   $\partial_g(ff^\pr)=\partial_g(f)R_g(f^\pr)+f\partial(f^\pr)$.
   $\partial_g$ is fixed to be $R_g-Id$ up to a constant depending on $g$, which can be absorbed
   by rescaling $\xi^g$.
   A {\it differential calculus} $(\Omega(\Alg[G]),d)$ is $\Alg[G]$-tensor product $\bigotimes_{\Alg[G]}M$
   modulo equivalent relations generated by the requirement that the
   extension of $d$ over this tensor product satisfying nilpotent rule and graded Leibnitz
   rule. In \cite{F1}, Feng \etal proved that
   if $G$ is a direct product of subgroups $H_1$ by $H_2$, and $G^\pr\subset(G^\pr\cap H_1)\cup(G^\pr\cap H_2)$,
   then
   \eq\label{DP}
    \xi^{h_1}\otimes\xi^{h_2}+\xi^{h_2}\otimes\xi^{h_1}\cong 0,
    \forall h_1\in G^\pr\cap H_1, h_2\in G^\pr\cap H_2.
   \en
   An {\it involution} $\inv$ on $G$ is an order-2 bijection;
   it is pulled back to be an involution on $\Alg[G]$ as $(\inv^\ast
   f)(g)=\overline{f(\inv(g))}$. Only Those $\inv\in Aut(G)$ will
   be considered below.
   An involution on $\Omega(\Alg[G])$ is an antilinear antihomomorphism
   which commutes skewly with $d$.
   {\it Main involution} $\ast$ on $\Alg([G])$ is defined by $\inv=Id$;
   the necessary condition of extensiblility of $\ast$ to an involution on $\Omega(\Alg[G])$
   is that $(\xi^g)^\ast=-\xi^{g^{-1}}$ and that
   $g\in G^\pr\Leftrightarrow g^{-1}\in G^\pr$.
   If $G$ is abelian, the structure of $G$ is classified by the
   following theorem \cite{R1}
   \begin{theorem}\label{Th1}
    An abelian group is finitely generated if and only if it is a
    direct product of finitely many cyclic groups of infinite or prime-power
    orders.
    \footnote{In mathematical literature, direct product of
    abelian groups is also referred as direct sum.}
   \end{theorem}
   $G$ will be assumed to be abelian henceforth.
   In this circumstance,
   {\it parity} $r$ on $G$ which is defined by $r(g)=-g$ can be extended to be an involution
   over $\Omega(\Alg[G])$ with any $G^\pr$. One can check that $r^\ast(\xi^g)=-\xi^g$.
  \section{Canonical Differential Calculi}\label{secC}
   As being remarked under Eq.(\ref{FN}), $\Alg[G]$ is a
   set-theoretical algebraic object, washing all information of
   $G$ out, and the structure of $G$ is {\it partly} recovered by
   Eq.(\ref{FN}) with $R_g\in End_\complex(\Alg[G]), g\in G^\pr$.
   Therefore, the following definition makes good sense.
   \begin{definition}
    A differential calculus is canonical, if 1) $G^\pr$ generates $G$,
    2) there exists an involution which is an extension of one $\inv^\ast$,
    3) $G^\pr$ is the minimum within all subsets of $G$ which satisfy 1) and 2).
   \end{definition}
   Follow Theorem \ref{Th1}, $G$ can be uniquely decomposed as a direct product of $\intg^{n_I}$ with a torsion subgroup
   $T=\bigotimes_{p,n_{pi},k_{pi}} (\intg_{p^{n_{pi}}})^{k_{pi}}$ where $p$ belongs to a finite set of prime numbers, $n_I,
   n_{pi},k_{pi}\in\nat$. From group theory, there exists
   a {\it basis} for $G$ which is the minimum generator set of $G$ and any element in $G$ is $\intg$-linear combination
   of this basis; more specifically,
   the basis is formed just by generators of each factor group in the decomposition of $G$, written as
   $\sigma_\alpha$ for $\intg_2$ factors, $T_s$ for $\intg$ factors, and $t_a$ for
   others. It is easy to check the following statement.
   \begin{proposition}
    If $G^\pr$ is taken to be the basis of $G$ and involution is taken to be parity, then the correspondent
    calculus is canonical.
   \end{proposition}
   \begin{corollary}\label{Coro1}
    If $G^\pr$ is taken to be the basis of $G$ and involution is taken to be
    parity, then     \\
    1) $\xi^g$ from different factor groups anticommute.\\
    2)
    \eq\label{MC1}
     d(\xi^{\sigma_\alpha})-2\xi^{\sigma_\alpha}\ot\xi^{\sigma_\alpha}\cong 0,
     d(\xi^{T_s})\cong 0, d(\xi^{t_a})\cong 0,
     \xi^{T_s}\ot\xi^{T_s}\cong 0,
     \xi^{t_a}\ot\xi^{t_a}\cong 0.
    \en
   \end{corollary}
   \prf
    1) is inferred by Eq.(\ref{DP}). To show 2) only $\intg_2$,
    $\intg$ and $\intg_n,n\geq 3$ need to be considered, and
    Eq.(\ref{MC1}) is inferred by $d(d(f))=0$.
   \endprf\\
   The following statement is also obvious.
   \begin{proposition}
    If $G^\pr=\{\sigma_\alpha, T_s, -T_s, t_a, -t_a: \mbox{for all }\alpha,s,a\}$ and the involution is taken to be the main
    involution, then the correspondent calculus is canonical.
   \end{proposition}
   The corollary below can be verified in the same way as that for
   Corollary \ref{Coro1}.
   \begin{corollary}
    1) $\xi^g$ from different factor groups anticommute.\\
    2) $d(\xi^{\sigma_\alpha})-2\xi^{\sigma_\alpha}\ot\xi^{\sigma_\alpha}\cong
    0$; $\xi^\tau\ot\xi^\tau\cong 0\cong\xi^{-\tau}\ot\xi^{-\tau}$,
    $d(\xi^\tau)-\{\xi^\tau, \xi^{-\tau}\}_\ot\cong 0$.
   \end{corollary}
  \section{Fermionic Representations of Calculi}\label{secF}
   Following Connes \cite{Co1},
   a {\it quantized calculus} over $\Alg[G]$ is a quadruple $(\Hs, \pi, F, \J)$
   where $\Hs$ is separable Hilbert space, $\pi$ is a faithful representation of $\Alg[G]$ by bounded operators on $\Hs$,
   $\J$ is representation of $\inv^\ast$ by $\pi(\inv^\ast(f))=\J\pi(f)^\dag \J^\dag, \forall f\in \Alg[G]$,
   hence $\J^2=Id$, moreover $\J$ is required to be unitary, and $F$ is a $\J$-selfadjoint operator
   \footnote{
    That $F$ is $\J$-selfadjoint is weaker than
    that $F$ is selfadjoint and commutes with $\J$, which is adopted in other literature.
   }
   which satisfies that $[F,\pi(f)]$ is compact for all $f$ in $\Alg[G]$
   and $F^2=Id$ up to a scalar.
   A quantized differential is defined to be
   $\hat{d}:\pi(\Alg[G])\rightarrow {\mathcal K}(\Hs), \pi(f)\mapsto i[F,\pi(f)]$
   where ${\mathcal K}(\Hs)$ is the algebra of compact operators on $\Hs$.
   A $p$-form in quantized calculus is of the form of linear combination of
   $i^p\pi(f_0)[F,\pi(f_1)][F,\pi(f_2)]\ldots[F,\pi(f_p)]$.
   $\hat{d}$ is extended to be a coboundary operator through graded adjoint action by $iF$.
   \begin{definition}
    A fermionic representation of a canonical calculus over $\Alg[G])$ is a quantized calculus over
    $\Alg[G]$ whose $\pi$ is extended to be an involutive differential representation
    of $\Omega(\Alg[G])$.
   \end{definition}
   The extension of $\pi$ is written as $\tld{\pi}$ with
   $\tld{\pi}(f)=\pi(f), \forall f\in \Alg[G]$.
   \begin{theorem}\label{THiff}
    A fermionic representation of a canonical calculus over $\Alg[G])$ exists,
    iff for an assignment of $\{\tld{\pi}(\xi^g): g\in G^\pr\}$ the following
    conditions are satisfied\\
    \eqa
    \label{iff1}
     \tld{\pi}(f\xi^g)=\pi(f)\tld{\pi}(\xi^g), \tld{\pi}(\xi^g
     f)=\tld{\pi}(\xi^g)\pi(f)\\
    \label{iff2}
     V:=iF-\sum_{g\in G^\pr}\tld{\pi}(\xi^g)\in \pi(\Alg[G])^\pr\\
    \label{iff3}
     \sum_{h,g-h\in G^\pr}\tld{\pi}(\xi^h)\tld{\pi}(\xi^{g-h})=0,
     g\in G\setminus(\{e\}\cup G^\pr)\\
    \label{iff4}
     \sum_{h,g-h\in G^\pr}\tld{\pi}(\xi^h)\tld{\pi}(\xi^{g-h})+\{V,\tld{\pi}(\xi^g)\}=0,
     g\in G^\pr\\
    \label{iff5}
     \tld{\pi}(\inv^\ast(\xi^g))=\J\tld{\pi}(\xi^g)^\dag\J^\dag
    \ena
    where $\pi(\Alg[G])^\pr$ is commutant of $\pi(\Alg[G])$ in
    $\comp(\Hs)$.
   \end{theorem}
   \prf
    {\it Necessity}: if $\tld{\pi}$ exists, then it must be a module
    homomorphism from $M$ to $\comp(\Hs)$, which implies
    Eq.(\ref{iff1}); Eq.(\ref{iff2}) is inferred by the commutative diagram
    $\hat{d}\circ\pi=\pi\circ d$;
    Eqs.(\ref{iff3})(\ref{iff4}) are implied by
    $\hat{d}\circ\pi\circ d=0$ which is the nilpotent rule of
    differential; Eq.(\ref{iff5}) follows that $\hat{\pi}$ is involutive
    representation.\\
    {\it Sufficiency}: Assume Eq.(\ref{iff1}) holds, then $\pi$ can be
    extended to be module homomorphism from $\bigotimes_{\Alg[G]}
    M$ to $\comp(\Hs)$; and Eq.(\ref{iff1}) implies that
    Eq.(\ref{FN}) is realized by
    $\tld{\pi}(\xi^g)\pi(f)=\pi(R_g(f))\tld{\pi}(\xi^g)$, with
    that $\tld{\pi}(d(f))=[\sum_{g\in
    G^\pr}\tld{\pi}(\xi^g),\pi(f)]$ following.
    Consequently, first order differential $(M,d)$ can be implemented by
    $\hat{d}\circ\pi=\pi\circ d$,
    thanks to Eq.(\ref{iff2}). High order representation of $d$ is
    realized by Eqs.(\ref{iff3})(\ref{iff4}) and the identity
    $ABC-(-)^{p+q}BCA=(AB-(-)^pBA)C+(-)^pB(AC-(-)^qCA)$, which
    guarantee the nilpotent rule and graded Leinitz rule of differential, together
    with that quotient conditions making tensor product of $M$ be
    $\Omega(\Alg[G])$ are contained in $ker(\tld{\pi})$. In the
    end, remember that $\inv^\ast$ is realized by $\J$, involutive homomorphism is extended
    from Eq.(\ref{iff5}) by antilinearity and antihomomorphic rule.
    That $F$ is $\J$-selfadjoint implies that skew-commutativity
    of involution and differential is realized.
   \endprf\\
   Fermionic representation exists for canonical calculus on $G$
   with main involution. In fact, introduce a set of fermionic creation and
   annihilation operators for each factor group of $G$ as
   $b^{\sigma_\alpha}, b^{T_s}, b^{T_s\dag}, b^{t_a},
   b^{t_a\dag}$ fulfilling anitcommutative relations
   \eqa\label{Cl1}
    \{b^{T_s},b^{T_r}\}=0, \{b^{T_s},b^{T_r\dag}\}=\dlt^{sr},
    \{b^{T_s},b^{t_b}\}=0, \{b^{T_s},b^{t_b\dag}\}=0,
    \{b^{T_s},b^{\sigma_\beta}\}=0,\\
   \label{Cl2}
    \{b^{T_s\dag},b^{T_r}\}=\dlt^{sr}, \{b^{T_s\dag},b^{T_r\dag}\}=0,
    \{b^{T_s\dag},b^{t_b}\}=0, \{b^{T_s\dag},b^{t_b\dag}\}=0,
    \{b^{T_s\dag},b^{\sigma_\beta}\}=0,\\
   \label{Cl3}
    \{b^{t_a},b^{T_r}\}=0, \{b^{t_a},b^{T_r\dag}\}=0,
    \{b^{t_a},b^{t_b}\}=0, \{b^{t_a},b^{t_b\dag}\}=\dlt^{ab},
    \{b^{t_a},b^{\sigma_\beta}\}=0,\\
   \label{Cl4}
    \{b^{t_a\dag},b^{T_r}\}=0, \{b^{t_a\dag},b^{T_r\dag}\}=0,
    \{b^{t_a\dag},b^{t_b}\}=\dlt^{ab}, \{b^{t_a\dag},b^{t_b\dag}\}=0,
    \{b^{t_a\dag},b^{\sigma_\beta}\}=0,\\
   \label{Cl5}
    \{b^{\sigma_\alpha},b^{T_r}\}=0, \{b^{\sigma_\alpha},b^{T_r\dag}\}=0,
    \{b^{\sigma_\alpha},b^{t_b}\}=0, \{b^{\sigma_\alpha},b^{t_b\dag}\}=0,
    \{b^{\sigma_\alpha},b^{\sigma_\beta}\}=2\dlt^{\alpha\beta}.
   \ena
   in which $b^{\sigma_\alpha}$ for $\intg_2$ factors
   are selfadjoint.
   Let $S$ be a representation of the above fermionic operators and
   $\Hs=l^2(G)\ot S$ with $\pi(f)=f\ot Id$.
   Define $\tld{\pi}(\xi^{\sigma_\alpha})=ib^{\sigma_\alpha}R_{\sigma_\alpha}$,
   $\tld{\pi}(\xi^{-T_s})=ib^{T_s}R_{-T_s}$, $\tld{\pi}(\xi^{T_s})=ib^{T_s\dag}R_{T_s}$,
   $\tld{\pi}(\xi^{-t_a})=ib^{t_a}R_{-t_a}$,
   $\tld{\pi}(\xi^{t_a})=ib^{t_a\dag}R_{t_a}$ and
   \eq\label{Fop}
    F=b^{\sigma_\alpha}R_{\sigma_\alpha}+b^{T_s}R_{-T_s}+b^{T_s\dag}R_{T_s}+b^{t_a}R_{-t_a}+b^{t_a\dag}R_{t_a},
   \en
   $\J=Id$. Then one can check that $F^\dag=F$, $F^2\sim Id$. Take Eq.(\ref{iff1}) as definition,
   Eq.(\ref{iff2}) is obvious with $V=0$, and
   Eqs.(\ref{iff3})(\ref{iff4}) are implied by Eqs.(\ref{Cl1})..(\ref{Cl5}).
   Eq.(\ref{iff5}) can be checked easily. Hence due to Theorem
   \ref{THiff}, above $(\Hs, \pi, \J, F)$ provides a fermionic
   representation of canonical calculus with main involution.  \\

   It is not so easy to construct fermionic representation for canonical calculus whose involution is taken to be
   parity.
   In fact, let $\J=r\ot Id$, then $\J\pi(f)^\dag\J^\dag=\pi(r^\ast(f))$.
   Introduce a set of Clifford generators $\gamma^{\sigma_\alpha}$, $\gamma^{T_s}$,
   $\gamma^{t_a}$ for each factor group of $G$, fulfilling that
   \[
    \{\gamma^{\sigma_\alpha},\gamma^{\sigma_\beta}\}=2\dlt^{\alpha\beta},
    \{\gamma^{\sigma_\alpha},\gamma^{T_r}\}=0,
    \{\gamma^{\sigma_\alpha},\gamma^{t_b}\}=0,
   \]
   \[
    \{\gamma^{T_s},\gamma^{\sigma_\beta}\}=0,
    \{\gamma^{T_s},\gamma^{T_r}\}=2\dlt^{sr},
    \{\gamma^{T_s},\gamma^{t_b}\}=0,
   \]
   \[
    \{\gamma^{t_a},\gamma^{\sigma_\beta}\}=0,
    \{\gamma^{t_a},\gamma^{T_r}\}=0,
    \{\gamma^{t_a},\gamma^{t_b}\}=2\dlt^{ab}.
   \]
   And define
   $F=\gamma^{\sigma_\alpha}R_{\sigma_\alpha}+\gamma^{T_s}R_{T_s}+\gamma^{t_a}R_{t_a}$.
   Then
   Eqs.(\ref{iff1})(\ref{iff2})(\ref{iff4})(\ref{iff5}) are valid.
   However, Eq.(\ref{iff3}) can not be realized unless requiring the product of conjunct
   $\gamma^{T_s}$, $\gamma^{t_a}$ to be wedge product. This will also
   guarantee that $F^2\sim Id$ in the above sense. In this
   circumstance, a weak form of fermionic representation is found
   out.
  \section{Discussions}\label{secA}
   When $G$ is taken to be $\intg^d$, a nontrivial correspondence between $F$ in Eq.(\ref{Fop})
   and staggered Dirac operator in lattice field theory was shown in
   \cite{DS1}. On the contrary, Eq.(\ref{Fop}) can be understood
   as a generalized staggered Dirac operator on FGAGs.\\

   Structures for non-FGAGs are far less clear in group theory, so the term
   canonical calculus for these groups are very blurred. Some specific choices have been explored in
   \cite{B1}\cite{Mj}.\\

  {\bf Acknowledgements}\\
    This work was supported by Climb-Up (Pan Deng) Project of
    Department of Science and Technology in China, Chinese
    National Science Foundation and Doctoral Programme Foundation
    of Institution of Higher Education in China.
  
 \end{document}